# Internet Appendix for "Sequential Bargaining Based Incentive Mechanism for Collaborative Internet Access"

Kübra Uludağ, Sinan Emre Taşçi and Ömer Korçak

*Abstract*— This document is an Internet Appendix of paper entitled "Sequential Bargaining Based Incentive Mechanism for Collaborative Internet Access". It includes information about LTE signal metrics, results of idle state experiments, and linear regression assumptions of the models presented in the related paper.

## APPENDIX A
## LTE SIGNAL METRICS

In LTE networks, radio resource management is performed according to Reference Signal Received Power (RSRP), Reference Signal Received Quality (RSRQ) and Signal to Interference plus Noise Ratio (SINR). These metrics provide information about signal strength and channel quality. RSRP is a cell-specific metric indicating signal strength that is primarily used for cell selection and handover. RSRP measurement does not include noise or interference from neighboring or serving cell [1], [2]. It can be affected by multipath fading and so RSRP reading can be seen low while the actual throughput may be high [3]. The reporting range of RSRP is defined from -140 dBm to -44 dBm with 1 dB resolution. RSRQ is a cell-specific signal quality metric that provides extra information to determine when to perform a handover. It provides a ranking among different candidate cells as regards their signal quality. It can be used together with the RSRP in making cell reselection and handover decision when the RSRP measurements are inadequate to make reliable decisions. The reporting range of RSRQ is defined from -19.5 dB to -3 with 0.5 dB resolution [4]. SINR is a measure of signal quality that can be used to evaluate effect of interference upon the link. It is not defined in 3GPP specifications; it is specified by user equipment vendors. [5].

## APPENDIX B
## IDLE STATE EXPERIMENTS

In order to calculate the net energy consumed by a smartphone during data downloading, it is necessary to measure the energy consumed by the device in the idle state and subtract it from the total consumed energy. The idle state means that no application is running on the background or foreground, except our application. We performed some idle running tests and during these tests, WiFi, Bluetooth, hotspot and mobile data were turned off and no applications were running on the background or foreground except our measurement application. The idle running tests on smartphones are basically divided into two groups depending on the screen status: Screen On and Screen Off. If the smartphone screen is off, the brightness level is not important. However, if the screen is on, the brightness level becomes important in terms of the power consumption. Some tests were done to see the effect of the smartphone's screen state (on / off) and screen brightness level on power consumption. The power values spent in the idle state given in Table B.1 are used to calculate the net energy consumed on a smartphone during the data download. While the calculation of energy consumption during the data download, the power spent in the idle state is subtracted from the instant total spent power. Then the net energy is calculated by using these net power values.

TABLE B.1
POWER CONSUMPTION TESTS IN IDLE STATE

| Display | Samsung Galaxy SM-J700F/H | | LG G3 | | Asus Zenfone 2 | |
|---|---|---|---|---|---|---|
| | Brightness | Avg. Power (W) | Brightness | Avg. Power (W) | Brightness | Avg. Power (W) |
| Off | - | 0.446 | - | 0.092 | - | 0.294 |
| On | 5[a] | 0.802 | 50[a] | 0.449 | 15[a] | 0.690 |
| On | 55 | 0.919 | 80 | 0.481 | 44 | 0.765 |
| On | 105 | 1.086 | 110 | 0.509 | 84 | 0.884 |
| On | 155 | 1.320 | 170 | 0.586 | 124 | 0.987 |
| On | 205 | 1.468 | 200 | 0.688 | 164 | 1.105 |
| On | 255[b] | 2.025 | 255[b] | 1.062 | 204[b] | 1.410 |

[a]*Min*, [b]*Max*

## APPENDIX C
## LINEAR REGRESSION ASSUMPTIONS

**Linearity.** There must be a linear relationship between a dependent variable (DV) and independents variables (IVs). More specify, dependent variable must be a linear function of an independent variable(s). If the linearity assumption is violated, the estimates such as regression coefficients, standard errors and test of statistical significance may be biased. The violations in the linearity assumption can be detected with LOWESS (Locally Weighted Scatterplot Smoothing) curve [6] that is a method of creating a smooth line that represents the relationship between independent variable(s) and dependent variable in a scatterplot. The LOWESS line looks linear (or rough linear) if the relationship between the dependent variable and independent variable(s) is linear (or approximately linear).

**Independence of Residuals (IoR).** There should not be a meaningful relationship between the residuals. If it is present autocorrelation (serial correlation) between the residuals, the standard errors and significance tests will be affected by that and they will not be accurate. The violations of the assumption can be diagnosed by Durbin-Watson test [7]. Durbin-Watson value lines between 0 and 4. The best value is 2 which means that there is no autocorrelation between the residuals. The values under 1 or more than 3 are a definite cause for concern [8].

**Constant Variance (Homoscedasticity).** The residuals should be constant variance (CV) for all values of IVs. This means that the variance of the dependent variable should not change depending on values of IVs. When there exists changing variance (heteroscedasticity) problem in a regression model, the estimates of standard errors and hence significance tests and confidence intervals will be incorrect. Homoscedasticity can be visually checked via scatterplots. Ideally, if a plot of standardizes residuals by standardized predicted values is randomly scattered around zero line, it can be said that the homoscedasticity assumption is met. Another way of checking the constant variance assumption is to apply Levene's Test [9]. In the Levene's, the residuals are initially divided into two or more groups, and a value is calculated to assess the equality of variances for the groups. If the significance value (p-value) of Levene's test is greater than 0.05, equal variance assumption is satisfied.

**Normality.** The residuals have normal distribution. Violations of the normality assumption do not cause biased estimation of regression coefficients. However, they can affect the estimates of the significance tests and confidence intervals depending on the sample size. The normality of the residuals can be detected statistically by Jarque Bera test [10]. According to this test, if the significance value (p-value) is greater than 0.05, the assumption is satisfied.

**Multicollinearity.** IVs are not highly correlated with each other. Multicollinearity is analyzed through collinearity statistics such as Condition Index (CI). CI is a statistic for diagnosing of the multicollinearity. If the CI is above 30, it indicates highly severe problems of multicollinearity. Some authors have proposed values of 15 or 20 as a threshold value for the CI [6].

In a linear regression model, independent variables can affect each other. Therefore, the interaction of independent variables should be examined and if there is an interaction between the variables (namely if the significance value of the interaction term is less than 0.05), the interaction term that is the multiplication of two independent variables should be included in the regression model.

## APPENDIX D
## ASSUMPTIONS OF STANDALONE AND GATEWAY MODELS

Linear regression assumption results of the standalone models are given in Table D.1.

TABLE D.1 STANDALONE MODELS ASSUMPTIONS

| Model | Assumptions | | | | | | |
|---|---|---|---|---|---|---|---|
| | Linearity (Lowess Curve) | IoR* (D-W Test) | CV** (Levene's Test) | | Normality (Jarque Bera) | | Multi-collinearity (CI) |
| | | | Stat | Sig. | Stat | Sig. | |
| 1.3 | Linear | 1.44 | 0.17 | 0.68 | 1.76 | 0.42 | 2.12 |
| 1.4 | Linear | 1.50 | 0.31 | 0.58 | 1.18 | 0.56 | 2.33 |
| 2.4 | Approx. Linear | 1.48 | 0.05 | 0.82 | 3.39 | 0.18 | 2.36 |
| 2.5 | Approx. Linear | 1.49 | 0.04 | 0.85 | 3.69 | 0.16 | 2.40 |
| 3.2 | Linear | 1.09 | 1.82 | 0.18 | 4.13 | 0.13 | 1.45 |
| 3.3 | Linear | 1.13 | 1.46 | 0.23 | 4.05 | 0.13 | 1.53 |

*Independece of Residuals, **Constant Variance

The results of the assumption of linearity are shown in the Figures D.1a., D.1.b, D.1.c, D.1.d, D.1.e, and D.1.f.

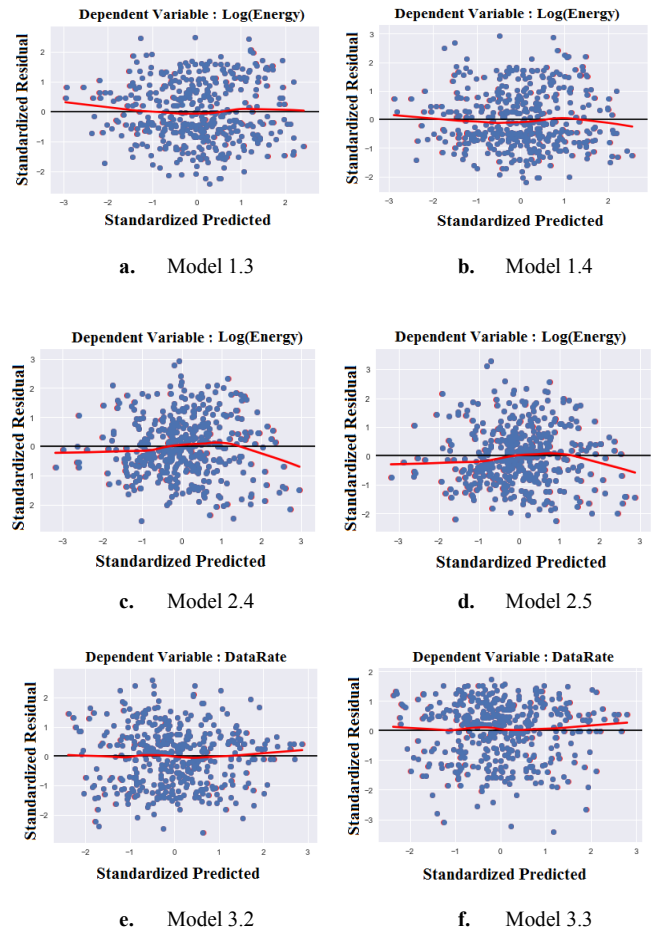

a. Model 1.3  b. Model 1.4
c. Model 2.4  d. Model 2.5
e. Model 3.2  f. Model 3.3

Fig. D.1 Scatterplots of the standalone energy consumption and data rate models

Linear regression assumption results of the gateway models are given in Table D.2.

TABLE D.2
MODEL ASSUMPTIONS FOR THE GATEWAY ENERGY CONSUMPTION AND DATA RATE

| Model | Assumptions | | | | | | |
|---|---|---|---|---|---|---|---|
| | Linearity (Lowess Curve) | IoR (D-W Test) | CV (Levene's Test) | | Normality (Jarque Bera) | | Multi-collinearity (CI) |
| | | | Stat | Sig. | Stat | Sig. | |
| Log ($E_g^{oc}$) | Approx. Linear | 1.22 | 0.00 | 0.97 | 5.33 | 0.01 | 1.06 |
| Data Rate ($R_g^{oc}$) | Approx. Linear | 1.84 | 2.65 | 0.11 | 0.91 | 0.64 | 1.04 |

The results of the assumption of linearity are shown in the Figures D.2.a and D.2.b.

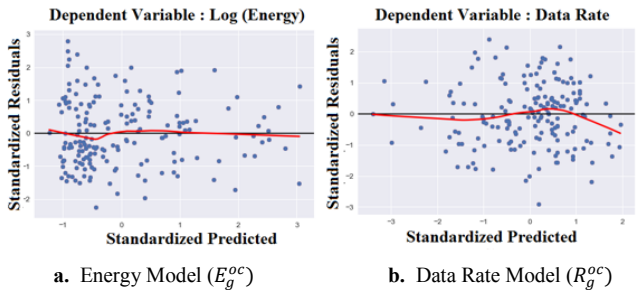

    **a.** Energy Model ($E_g^{oc}$)      **b.** Data Rate Model ($R_g^{oc}$)

Fig. D.2 Scatterplots of the gateway energy consumption and data rate models